\documentclass[]{spie}  

\usepackage{amsmath,amsfonts,amssymb}
\usepackage{graphicx}
\usepackage[colorlinks=true, allcolors=blue]{hyperref}
\usepackage{cite}
\newcommand{\brg}{Br$\gamma$} 
\usepackage{sidecap}

\title{The innermost astronomical units of protoplanetary disks}

\author[a]{Jacques Kluska}
\author[b]{Rebeca Garc\'ia L\'opez}
\author[c]{Myriam Benisty}
\affil[a]{University of Exeter, School of Physics, Stocker Road, Exeter, EX4 4QL, UK}
\affil[b]{Dublin Institute for Advanced Studies, 31 Fitzwilliam Place, Dublin 2, Ireland}
\affil[c]{Univ. Grenoble Alpes, IPAG, F-38000 Grenoble, France; CNRS, IPAG, F-38000 Grenoble, France}

\authorinfo{Further author information: E-mail: jkluska@astro.ex.ac.uk}

\begin{document} 
\maketitle

\begin{abstract}
Circumstellar disks around young stars are the birthsites of planets. It is thus fundamental to study the disks in which they form, their structure and the physical conditions therein. The first astronomical unit is of great interest because this is where the terrestrial-planets form and the angular momentum is controled via mass-loss through winds/jets. With its milli-arcsecond resolution, optical interferometry is the only technic able to spatially resolve the first few astronomical units of the disk. In this review, we will present a broad overview of studies of young stellar objects with interferometry, and discuss prospects for the future. 
\end{abstract}

\section{Introduction}

Protoplanetary disks are believed to be the birthplace of planets, and of particular interest is the innermost AUs where terrestrial planets are formed. This region is therefore of prime interest to study the initial conditions of planet formation. It is also within the first AU that material is accreted onto the star and that stellar and disk winds take place.  

For a long time these regions were mainly studied through spectro-photometry as their angular extension was unreachable by monolithic telescopes. The dust emits in the continuum and leads to a significant excess emission to the photosphere in the spectral energy distribution (SED). However, model fitting of the SED is highly degenerate and only spatially resolved observations can disentangle between the different models and constrain the morphology and kinematics of the disk material.  


Optical interferometry overcomes the resolution limit and provides spatial information of the inner parts of the protoplanetary disks at milli-arcsecond scales.
Early interferometric observations found that the near-infrared sizes correlate with the stellar luminosity, in agreement with the fact that the NIR emission traces hot dust located at the dust sublimation front. With the advent of mutiple beam combiners, more detailed modelling and imaging studies of the dust rim could be achieved. Spectro-interferometric observations of the Hydrogen lines shed light on accretion and ejection mechanisms occurring very close to the star. Finally, the location and size of gaps was inferred from near-infrared and mid-infrared interferometric observations. 

The majority of these findings are already described in recent reviews\cite{DM2010,Kraus2015}.
In this paper we will focus on the new discoveries that were not discussed in previous reviews.
Our review is divided in three sections.
Section \ref{sec:Subl} covers the dust sublimation morphology, Section \ref{sec:TD} will cover the constraints on gaps and planetary companions in (pre-)transitional disks and Section \ref{sec:lines} described the recent advances in spectro-interferometry on the dynamics of the very inner regions.
Finally, we will conclude with Section \ref{sec:ccl}.

\section{Morphology of the sublimation front}
\label{sec:Subl}

 \begin{figure} [th]
   \begin{center}
   \begin{tabular}{c} 
  \includegraphics[height=8cm]{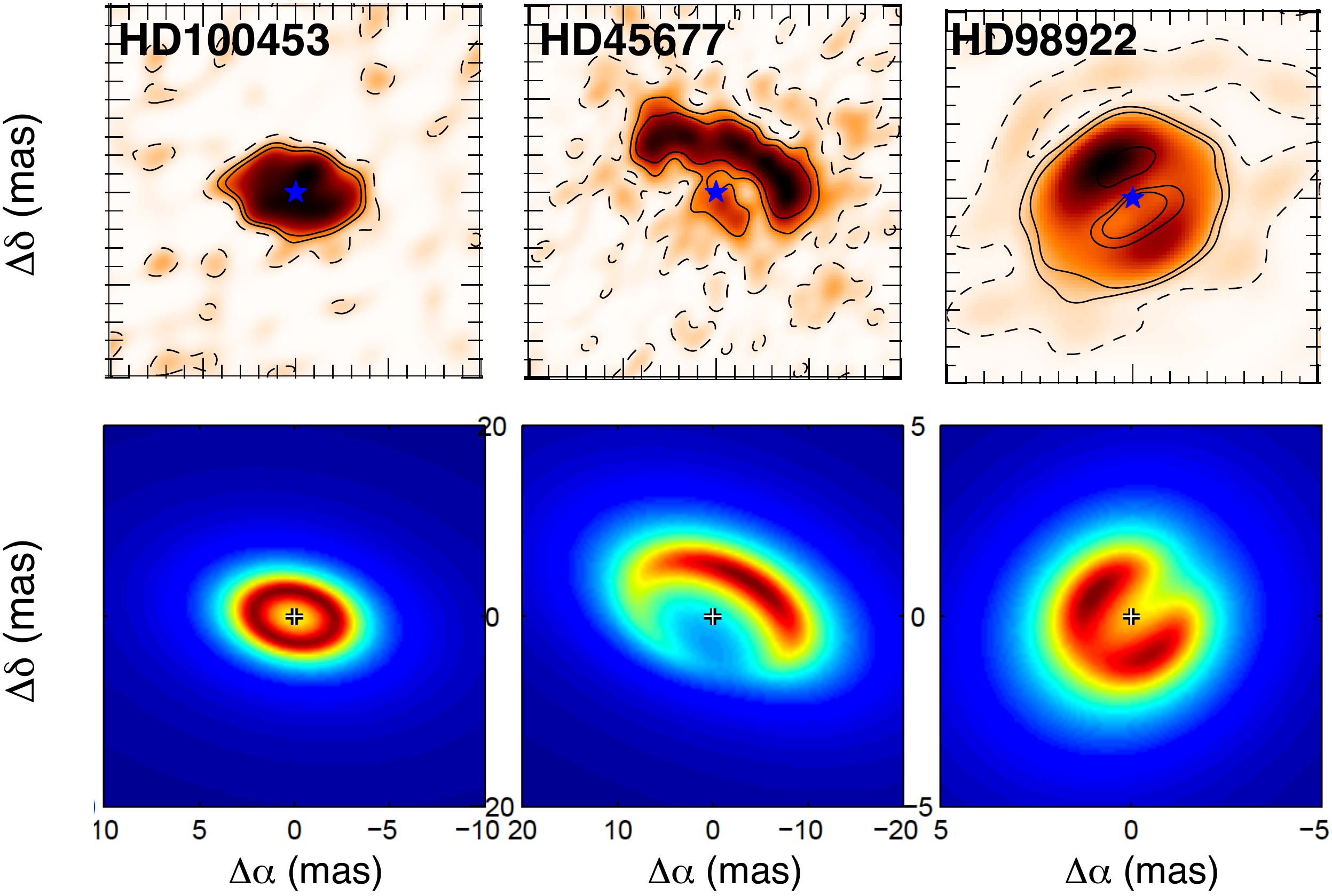}
   \end{tabular}
   \end{center}
   \caption 
   { \label{fig:1} Relation between image reconstructions (first row) and model fitting (second row) on data from the PIONIER/VLTI survey of Herbig Ae/Be stars. For some objects the reconstructed image suggest more complex structures of the inner rim. These images motivate the use of models with higher azimuthal modulation degree to reproduce the closure phases.}
   \end{figure} 


\subsection{Herbig stars}
  
A large survey was conducted with PIONIER on 51 Herbig Ae/Be stars, intermediate mass (2-8 M$_\odot$) young stars.
PIONIER\cite{LeBouquin2011} is a near-infrared interferometric instrument, that combines 4 telescopes and that is installed at the Very Large Telescope Interferometer (VLTI).
Previous surveys have contributed to build the size of near-infrared emission vs. stellar (and accretion) luminosity relation.
The aim of the survey was to constrain the 3D structure of the inner disk rim, to constrain the dust properties and detect additional components contributing to the emission. 

\subsubsection{Statistical constrains}

One part of the survey was to derive statistical constraints on the inner disk rim (Lazareff et al. in prep.).
We will summarize here the main results.

{\bf A hot rim.} Spectrally dispersed continuum interferometric observations have information on the spectral slope of the emission from the circumstellar  environment\cite{Kluska2014}. Both interferometric measurements and photometric observations lead to a temperature estimate of 1800K, higher that the typical sublimation temperature for silicate dust grains ($\sim$1500K). 
These high temperatures can be explained by considering refractory dust species, such as Carbon grains in the sublimation front. An alternative scenario is the presence of hot gas inside the dust sublimation radius, that would contribute to the excess emission. Similar conclusion was reached by other studies\cite{Kama2009,Carmona2014}.

{\bf A smooth rim.} The near-infrared emission has a large radial extension and is rather smooth, in contrast with predictions of models with a sharp puffed-up rim \cite{Isella2005}.
This likely indicates that various grains populations, with slightly different sublimation temperatures, are present (dust segregation \cite{THM2007,Kama2009,Benisty2010,McClure2013}). 

{\bf A rather symmetric rim.} The closure phase measurements enable the detection of asymmetric signals coming from the disk. The detected  asymmetries are in majority consistent with radiative transfer effects due to disk inclination.
There is one outlier, MWC158, that shows a strong variability in its environment morphology\cite{Kluska2016}.
This effect can be due to additional sources of asymmetry located in the disk. 

A significant scatter is found in the the size-luminosity relation, although the global shape of this diagram is similar to previous size luminosity diagrams\cite{Monnier2002,Monnier2005}. 
This scatter can be explained by different inner rim models such as narrow or extended rims and with different particle sizes. 
This points out towards a large panel of inner rims within the Herbig Ae/Be stars.

\subsubsection{Imaging}
Imaging capabilities of optical interferometers significantly increased in the last years.
A specific imaging algorithm was developed (Semi-Parametric Approach for reconstruction of Chromatic Objects, SPARCO \cite{Kluska2014}).
In a simple version it incorporates a parametric model of the central star (the parametric model can therefore be more complex: binary, extended flux...).
Moreover the difference in spectral behaviours between the hot star and its colder environment is taken into account. 

The images of young stellar objects observed with PIONIER/VLTI reveal complex asymmetric structures (Kluska et al. in prep.).
The SPARCO imaging approach was applied to the Herbig AeBe survey dataset.
The images do not reveal narrow rings for most of the stars confirming the claims of the statistical analysis.
The orientations are also in agreement with what is found with model fitting.
Thanks to the closure phases it is also possible to probe the asymmetry of some objects.
They can be then checked to see if they are consistent with inclination effects.
This is the case for most objects.
Nevertheless the revealed asymmetries seem to be more complex that just inclination effects, confirming the findings with the statistical analysis.

For some objects the complex structure seen by imaging led to fitting the data with more complex structure (ex. higher degrees of azimuthal modulation, see Fig.\ref{fig:1}).
These more complex model were indeed more successful to fit the data.
Image reconstruction in optical interferometry are becoming mature and can more and more guide the analysis of objects showing complex structures unreachable by model fitting.

The areas targeted by optical interferometry can evolve on a time scale of weeks.
Quick changes in the object circumstellar morphology can smear out the image\cite{Kluska2016}.

\subsection{T Tauri stars}

T Tauri stars appear over-sized on the size luminosity diagram compared to the prediction from dust sublimation models\cite{Eisner2007}.
A possible explanation was the contribution from scattered light. Stellar light scattered on the disk surface can  make the object appear larger in interferometric observations\cite{Pinte2008}. 
A survey\cite{Anthonioz2015} of these stars was conducted with PIONIER.
More than half of the objects show a drop a short baselines.
This drop is successfully modelled by stellar light scattered on the inner parts of the disks.
These objects follow the size luminosity relation as expected when the scattered light is taken into account. 

\section{Evidence for gaps in the inner astronomical units}
\label{sec:TD}

Spatially resolved observations are complementing SED classification scheme for Herbig stars.
From photometric observations, a first classification was made\cite{Meeus2001}.
Objects with a double-peaked SED are classified as group I objects whereas objects with a single peaked SED are classified as group II.
A first interpretation of this classification was that group I sources are flared disks in opposition to flat disks group II objects\cite{Meeus2001}.
A recent interpretation claims that group I objects are gapped transitional disks and that no gaps appear in group II disks\cite{Maaskant2013}.

The PIONIER survey dataset points at an extended emission (over-resolved by the interferometer) for group I sources. It can be interpreted as stellar light scattered from the inner rim of the outer disk or by quantum heated particles located in the gap.

Mid infrared interferometry is an efficient technique to detect gapped disk structures at a few astronomical units scales.
A study of the archival MIDI/VLTI data was conducted\cite{Menu2015}.
Grids of radiative transfer models confirm that the majority of group I sources have gaps. Moreover several group II sources are not compatible with a model of a continuous disk suggesting a gap in the inner parts. Nevertheless, the size of this gap is smaller that in group I objects.
This study suggests a new evolutionary scheme for intermediate mass stars. Flat and flared disks could evolve separately from a common ancestor or gapped flat disks are the progenitors of transitional group I disks.

Another study in the mid-infrared was conducted with nulling interferometry technique\cite{MG2016}. 
A two layered disk structure\cite{McClure2013} can reproduce both SED and interferometric datasets for the majority of the objects. This points out towards gapped structure or more complex layered disks.
Other more detailed studies combining various interferometric datasets (near-infrared, mid-infrared, aperture masking) were used to constrain complex gapped disk structures around individual objects\cite{Benisty2010, Tatulli2011,Cieza2013,Kraus2013, Carmona2014,Matter2014, Matter2016}.

Proto-planets responsible for gaps can potentially be detected via the aperture masking technique
\cite{Huelamo2011,Olofsson2013,Kraus2013,Sallum2015}
Interestingly the discovery of a protoplanet in formation around LkCa15\cite{KrausIreland2012} was recently confirmed by direct imaging observations\cite{Sallum2015Nat} showing the complementarity between the different techniques.

\section{Accretion/wind on au-scale}
\label{sec:lines}

 \begin{SCfigure}
   \centering
   \caption[example] 
   { \label{fig:2} {\bf Left:} Intensity distributions of the disk-wind model best reproducing AMBER-HR observations of the Herbig AeBe star HD98922 (see {\emph Right}). The panels represent the intensity distribution of the continuum disk plus the disk wind at the centre of the Br$\gamma$ (v=0\,km/s) and at four other velocities \cite{CarattioGaratti15}. 
{\bf Right:} AMBER-HR observations of HD98922 (grey) and results from the disk-wind model (coloured) presented in {\emph Left}, showing the dependence of the interferometric observables on wavelength across the Br$\gamma$ line. }
  \includegraphics[height=9cm]{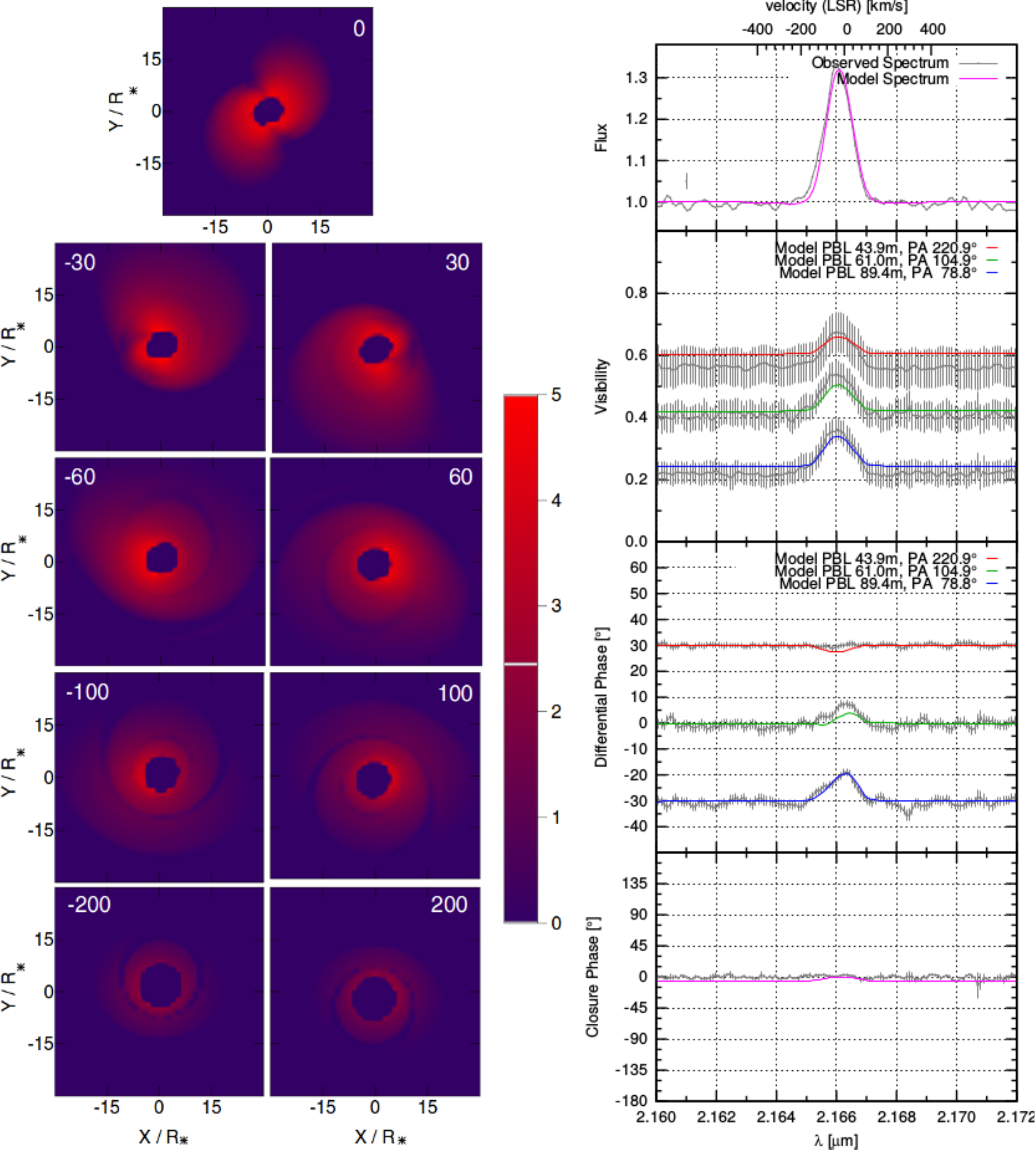}
   \end{SCfigure} 


One of the best studied tracers of the hot gas at sub-au scales is the HI \brg\ line. This line has been focussing most of the attention of near-IR spectro-interferometric studies, as it is the only tracer of the physical processes taking place in the inner gaseous disc that can be currently spatially resolved at near-IR wavelengths. 

In the context of young stellar objects (YSOs), the origin of this line is still uncertain. Near-IR interferometers have revealed that, in 
most cases, the \brg\ line originates from a very compact region (typical size of a few tenths of mas \cite{Kraus08,Eisner14}), well inside the dust 
sublimation radius. 
Within this region the accretion and ejection processes take place and the dust particles evaporate, giving rise to a very complex environment. Therefore, several physical processes and regions can 
contribute to the \brg\ emission: the magnetospheric accretion region, the wind emission, and the inner gaseous disk and hot
evaporative disk atmosphere.
Because of this complex scenario, a simultaneous physical modelling of both interferometric observables (visibilities, differential phases, and closure phases) and line profiles (ideally at high-spectral resolution) is crucial to discern the contribution of each region to the line emission.

Although the number of studied objects is still limited, this approach has now been applied to several Herbig AeBe stars yielding interesting results (see \cite{Weigelt11, Mendigutia15, Ellerbroek15, GarciaLopez15, CarattioGaratti15, GarciaLopez16, Kurosawa16} for more details on each study). 
In general, extended emission from a magneto-centrifugally driven wind launched from the inner regions of the gaseous disk ($\lesssim$0.1\,au) seems the best mechanism to simultaneously reproduce the \brg\ line profile and intensity, as well as the interferometric observables. 

Radiative transfer disk wind models have, however, problems reproducing the single-peaked line profiles observed in some objects at high inclination angles (see for instance \cite{GarciaLopez16}). The same problem is observed when considering that the \brg\ line is emitted in the surface layers of the disk \cite{Mendigutia15, Ellerbroek15}, since both the disk and the base of the disk wind are assumed to rotate at Keplerian velocity (see e.g. \cite{Kurosawa16}). In both cases,  emission from the magnetospheric accretion region, and/or the hot layers of a photoevaporative wind could contribute to the \brg\ emission, helping to produce a single-peaked line profile. However, recent studies seem to indicate that the contribution to the \brg\ emission of the magnetospheric region (too compact in Herbig AeBe stars, \cite{Tambovtseva14, GarciaLopez15, Kurosawa16, Tambovtseva16}), as well as the inner gaseous disk and/or hot photoevaporative wind (not dense enough to produce significant \brg\ emission, see \cite{Tambovtseva16}) is negligible in comparison with the contribution of the disk wind. 

Finally, the \brg\ inner launching radius of the disk wind seems to increase with the stellar luminosity (see Fig.\,9 in \cite{Kurosawa16}), with a much higher inner launching radius for the only late type Herbig Be star of the sample (MWC\,297). This result might indicate a difference in the strength of the wind and/or disk radiation pressure between early type Herbig AeBe stars and late type Herbig Be stars \cite{Kurosawa16}. Interestingly, the size of the \brg\ emitting region of the only massive YSO studied so far (IRAS 13481-6124) is much more extended ($\sim$3--6\,au \cite{CarattioGaratti16}), although well within the dust sublimation radius ($\sim$9\,au), than those measured in Herbig AeBe stars, hinting again at the presence of an extended disk wind and/or jet \cite{CarattioGaratti16}.   

Although promising, these results rely only on a small number of objects, observed with a limited coverage of the \emph{uv} plane. Further studies based on statistical samples with better \emph{uv} coverage as well as larger ranges of spectral types and evolutionary stages, are needed to finally clarify the origin of the \brg\ line and disentangle the kinematics and dynamics of the inner regions of protoplanetary disks.

\section{Conclusions}
\label{sec:ccl}

Optical interferometry is important to characterise the first astronomical units around the central star. 
The dust sublimation front is more extended for most of Herbig stars than initially predicted and image reconstruction process is now mature.
Studies of the \brg\ line revealed and characterised the disk wind taking place at and inside the dust sublimation radius.

Extended studies in both continuum and lines around young stellar object will help us understand the star disk interactions and the planet formation initial conditions.
The perspective of the new instruments that are or will be installed in the next few years is therefore very exciting.

\acknowledgments 
JK acknowledges support from an Marie Sklodowska-Curie CIG grant (Ref. 618910, PI: Stefan Kraus). RGL  is supported by Science Foundation of Ireland, grant 13/ERC/I2907.

\bibliography{report} 
\bibliographystyle{spiebib} 

\end{document}